\documentclass[tikz,10pt]{elsarticle} 
\usepackage{tikz}
\makeatletter
\def\ps@pprintTitle{%
  \let\@oddhead\@empty
  \let\@evenhead\@empty
  \def\@oddfoot{\reset@font\hfil\thepage\hfil}
  \let\@evenfoot\@oddfoot
}
\makeatother
\usetikzlibrary{shapes.geometric, arrows, positioning, chains}
\usepackage[utf8]{inputenc}
\usepackage{algorithm}
\usepackage{algpseudocode}
\usepackage{amsmath}

\usepackage{amsfonts}
\tikzstyle{startstop} = [rectangle, rounded corners, minimum width=3cm, minimum height=1cm,text centered, draw=black, fill=red!30]
\tikzstyle{process} = [rectangle, minimum width=3cm, minimum height=1cm, text centered, draw=black, fill=orange!30]
\tikzstyle{decision} = [diamond, minimum width=3cm, minimum height=1cm, text centered, draw=black, fill=green!30, aspect=2]
\tikzstyle{arrow} = [thick,->,>=stealth]

\usepackage{graphicx}

\usepackage{colortbl}
\usepackage[margin=2.0cm]{geometry}
\usepackage{hyperref} 
\usepackage{booktabs}
\usepackage{array}
\usepackage{longtable}
\usepackage{booktabs}
\usepackage{subcaption}
\usepackage{caption}
\definecolor{Highlight}{RGB}{255, 235, 59} 
\usepackage{tabularx}

\title{\textbf{Automated Traffic Incident Response Plans using Generative Artificial Intelligence: Part 1 - Building the Incident Response Benchmark }}
\author{Dr Artur Grigorev (\url{Artur.Grigorev@uts.edu.au}), University of Technology Sydney, Australia\\ Dr Khaled Saleh (\url{Khaled.Saleh@ieee.org}), University of Newcastle, Australia\\ Dr Jiwon Kim (\url{jiwon.kim@uq.edu.au}), The University of Queensland, Australia\\ Dr Adriana-Simona Mihaita (\url{Adriana-Simona.Mihaita@uts.edu.au}), University of Technology Sydney, Australia\\}
\date{}

\usepackage{listings}
\usepackage{xcolor}

\lstdefinestyle{mystyle}{
    basicstyle=\ttfamily\small,
    breaklines=true,
    frame=single,
    numbers=none,
    keywordstyle=\color{blue},
    commentstyle=\color{gray},
    stringstyle=\color{black},
    showstringspaces=false
}

\usepackage{float}
\begin{document}
\begin{abstract}

Traffic incidents remain a critical public safety concern worldwide, with Australia recording 1,300 road fatalities in 2024, which is the highest toll in 12 years. Similarly, the United States reports approximately 6 million crashes annually, raising significant challenges in terms of a fast reponse time and operational management. Traditional response protocols rely on human decision-making, which introduces potential inconsistencies and delays during critical moments when every minute impacts both safety outcomes and network performance. 

To address this issue, we propose a novel Incident Response Benchmark that uses generative artificial intelligence to automatically generate response plans for incoming traffic incidents. Our approach aims to significantly reduce incident resolution times by suggesting context-appropriate actions such as variable message sign deployment, lane closures, and emergency resource allocation adapted to specific incident characteristics. 
First, the proposed methodology uses real-world incident reports from the Performance Measurement System (PeMS) as training and evaluation data. We extract historically implemented actions from these reports and compare them against AI-generated response plans that suggest specific actions, such as lane closures, variable message sign announcements, and/or dispatching appropriate emergency resources. 
Second, model evaluations reveal that advanced generative AI models like GPT-4o and Grok 2 achieve superior alignment with expert solutions, demonstrated by minimized Hamming distances (averaging 2.96-2.98) and low weighted differences (approximately 0.27-0.28). Conversely, while Gemini 1.5 Pro records the lowest count of missed actions, its extremely high number of unnecessary actions (1547 compared to 225 for GPT-4o) indicates an over-triggering strategy that reduces the overall plan efficiency.

Our research demonstrates that carefully calibrated generative AI systems can provide valuable decision support for real-time traffic management by generating efficient incident response strategies that reduce resolution times and network disruption. Future research aims to evaluate various approaches to response enhancement including retrieval augmentation and guideline synthesis, to enhance the accuracy and adaptability of the proposed system.

\end{abstract}

\maketitle


\section{Introduction}

Traffic congestion stemming from road incidents represents a substantial and growing burden on urban environments and transportation networks. Beyond the immediate frustration of increased travel times, these incidents contribute significantly to elevated air pollution levels, pose direct risks to road user safety, and inflict considerable economic losses through wasted fuel and lost productivity \cite{cong}. Traffic Management Centers (TMCs) are tasked with mitigating these impacts, using a combination of automated systems (like sensor networks and adaptive signals) and manual processes driven by human operators \cite{de2017traffic,s16020157}. However, the sheer complexity of modern road infrastructure, coupled with the unpredictable nature and real-time evolution of traffic incidents, frequently overwhelms these traditional methods. Static algorithms and reactive measures often struggle to adapt dynamically, highlighting a critical need for more sophisticated, rapid, and data-driven decision-making capabilities in incident response \cite{mah}.


To address this, we propose the Incident Response Benchmark, a framework designed to evaluate automated traffic incident response solutions. This benchmark uses real-world incident records and documented human responses as a basis for comparison against Large Language Model (LLM) outputs inferred from PeMS dataset. By analyzing these incident reports, the benchmark systematically compares how closely AI-generated plans align with actual, on-the-ground strategies used by TMCs.

Despite its promise, the integration of AI into existing traffic management workflows remains complex for several reasons:

\begin{enumerate} 
\item \textbf{Complexity of Incident Situations}: Urban transportation infrastructures are inherently complex, featuring multiple intersections, layered roadways, and a high density of control infrastructure. This complexity amplifies the difficulty of accurately predicting and efficiently managing traffic disruptions when incidents occur.

 \item \textbf{Variety of Traffic Incidents}: Incidents range from minor breakages to large-scale car crashes, each requiring a unique response strategy.
 
  \item \textbf{Real-Time Decision Making}: Rapidly evolving incidents demand swift action, and any new information often necessitates revisiting and adapting the initially proposed plan.
\end{enumerate}

The Incident Response Benchmark encodes real-world incident responses as binary vectors, where each bit represents a specific operation or intervention performed during an event. By converting tangible actions, such as investigative tasks and containment procedures — into a uniform binary structure, the benchmark enables quantitative comparison of AI-generated strategies to empirical best practices. Central to this comparison is the Hamming distance, which measures the number of binary decisions in which an AI-produced vector differs from the human-derived ground truth. This metric not only quantifies the divergence, but also allows to highlight operations that the AI may consistently misjudge.

\section{Related Works}

AI-based traffic management approaches refer to the application of artificial intelligence algorithms, models, and systems to monitor, manage, and optimize the flow of vehicles in urban traffic networks. These approaches operate on multiple levels of automation, ranging from basic reactive methods to fully adaptive and predictive systems incorporating real-time data. At the foundational level, simple automation handles static, rule-based traffic operations, while more advanced approaches, such as intelligent transport systems (ITS), utilize sensor data, machine learning (ML), and optimization algorithms to dynamically control traffic. The integration of LLMs represents a paradigm shift in the development of traffic management systems, particularly in automating the generation of rapid incident response plans. LLMs excel in parsing unstructured text, can automate tasks like summarizing information, generating reports, classifying incident severity, and providing decision support to TIM operators \cite{mahmud2025integratingllmsitsrecent}. Ongoing research in the area of LLMs demonstrates their utility in various disaster management scenarios, including generating pre-incident contingency plans, real-time response strategies, and post-incident recovery plans \cite{xue2023, goecks2023}. However, these models often struggle with data integration (e.g. due to limited size of context window) and may lack the domain-specific knowledge required for accurate traffic incident management.


Multiple existing traffic management systems rely on rule-based techniques like Case-Based Reasoning (CBR) or Bayesian Theory \cite{ma2014automated, ravish2021intelligent}. While these approaches have shown some success, they often struggle with the scalability and adaptability required for real-time incident response. CBR systems are reliant on the quality and quantity of historical cases, which can limit their effectiveness in novel situations. Bayesian methods, while powerful, can become computationally expensive in real-time scenarios, especially with large datasets \cite{ma2014automated}.


The OpenAI's ChatGPT model previously has been explored for its capabilities in natural disaster management and demonstrates the applicability of LLM models in accelerating the disaster management process through stages such as pre-disaster, during-disaster, and post-disaster scenarios \cite{xue2023}. These models have been proven to be effective in generating rapid response plans and disseminating crucial information during the event, with inherent limitations concerning real-time data processing and professional accuracy. The model addresses the limitations in real-time data handling and expert-level accuracy in complex disaster management scenarios. Also, a recently proposed DisasterResponseGPT framework \cite{goecks2023} relies on LLMs to generate rapid action plans in disaster situations with limitations in handling visual and real-time data inputs. 


Through integrating Geographic Information System (GIS) capabilities with AI \cite{huang2007}, the efficiency of incident responses can be markedly improved by optimizing resource allocation and managing traffic disruptions in real-time scenarios. AI models are capable of facilitating quick and effective decision-making processes by processing and analyzing large volumes of data swiftly \cite{huang2007}. Works that tackle the integration of AI in traffic systems further affirm the potential of LLMs in transforming urban traffic management \cite{zhang2024}. The integration of real-time data, the need for continuous model training, and the development of robust frameworks to handle complex emergency situations are recurrent themes across studies. As indicated by safety analytics in transport and construction sectors, there remains a crucial gap in applying AI to dynamically interpret complex environmental data sets and interact seamlessly across diverse operational platforms. Real-time traffic incident management demands rapid assessment and response, which can be significantly optimized using Generative AI. There is a critical need for integrating real-time data with AI systems to manage incidents effectively on complex traffic networks \cite{huang2007integration}. The AI systems, however, are susceptible to hallucinations and biases \cite{salvagno2023artificial, sun2024ai, sahoo2024addressing}.

The integration of data-driven models in Emergency Response Management systems was previously extensively reviewed \cite{mukhopadhyay2023emergency} and highlighted the disconnect between traditional static models and the dynamic requirements of real-world emergency management. The authors proposed a transition towards more adaptive models, capable of real-time adjustments in response strategies based on live data inputs. Key challenges identified include the ethical implications which remain paramount as these technologies become more relevant in critical areas affecting public safety of predictive policing and the limitations of current technology in accurately modeling complex urban behaviors. Current study aims to address AI capabilities with a consideration of ethical implications, striving for transparency and reliability of AI-driven methodology in traffic incident management. The application of dynamic resource allocation has been demonstrated in a similar case of response planning: crime response, with strategies like predictive analytics and hotspot policing gaining prominence. The continuous-time Markov decision process (CTMDP) uses real-time crime data to optimize police patrol allocations, significantly improving response times in urban scenarios by minimizing the duration between crime occurrence and police arrival.

Generative AI has been used across various domains including cybersecurityto assist in incident response operations by deriving response policies and methods based on large amounts of past data \cite{10737408}. DynamicRouteGPT \cite{zhou2024dynamicroutegpt} introduces a sophisticated framework that uses a combination of causal inference, Bayesian methods, and Markov chain modeling to refine multi-vehicle navigation in real time; focuses on optimizing navigation paths through end-to-end integration of large-scale pretrained language models for explainable and adaptive decision-making, our work similarly emphasizes the importance of incorporating real-time data and advanced AI techniques to ensure that incident response plans are both efficient and aligned with human expertise.

Despite these advances, bridging the gap between static or narrowly scoped approaches and the dynamic requirements of real-world incident management remains challenging. Traditional models frequently lack the agility to adapt to evolving road conditions, changes in incident severity, and newly emerging data streams \cite{mukhopadhyay2023emergency}. Trust and transparency are additional concerns in any AI-driven emergency context \cite{tao2023}, underscoring the need for frameworks that not only make decisions but also clarify the reasoning behind them.
Ravish and Swamy's review discusses the integration of machine learning and computational intelligence in managing traffic congestion, highlighting significant advancements in Intelligent Transportation Systems (ITS). They stress the importance of real-time data processing and accident detection mechanisms for improving traffic flow and safety \cite{ravish2021intelligent}. The review identifies challenges such as the increasing number of vehicles and insufficient infrastructure expansion. Although machine learning and computational intelligence offer promising solutions, real-time data processing, system complexity, adaptability, and scalability, as well as accident detection, remain critical concerns for the advancement and expansion of traffic systems.

AI models demonstrate capabilities in performing complex text analysis, including the examination of accident reports to enhance the prediction of various aspects of traffic incidents \cite{zheng2023chatgpt}. However, the potential of Generative AI in generating immediate incident response plans remains largely unexplored. 
 
Overall, this study underscores the capacity of Generative AI to process complex information from the area of traffic management and generate actionable safety recommendations, illustrating the broader applicability of AI in managing hazards.


\section{Methodology}

\begin{figure}[H]
    \centering
    \includegraphics[width=\textwidth]{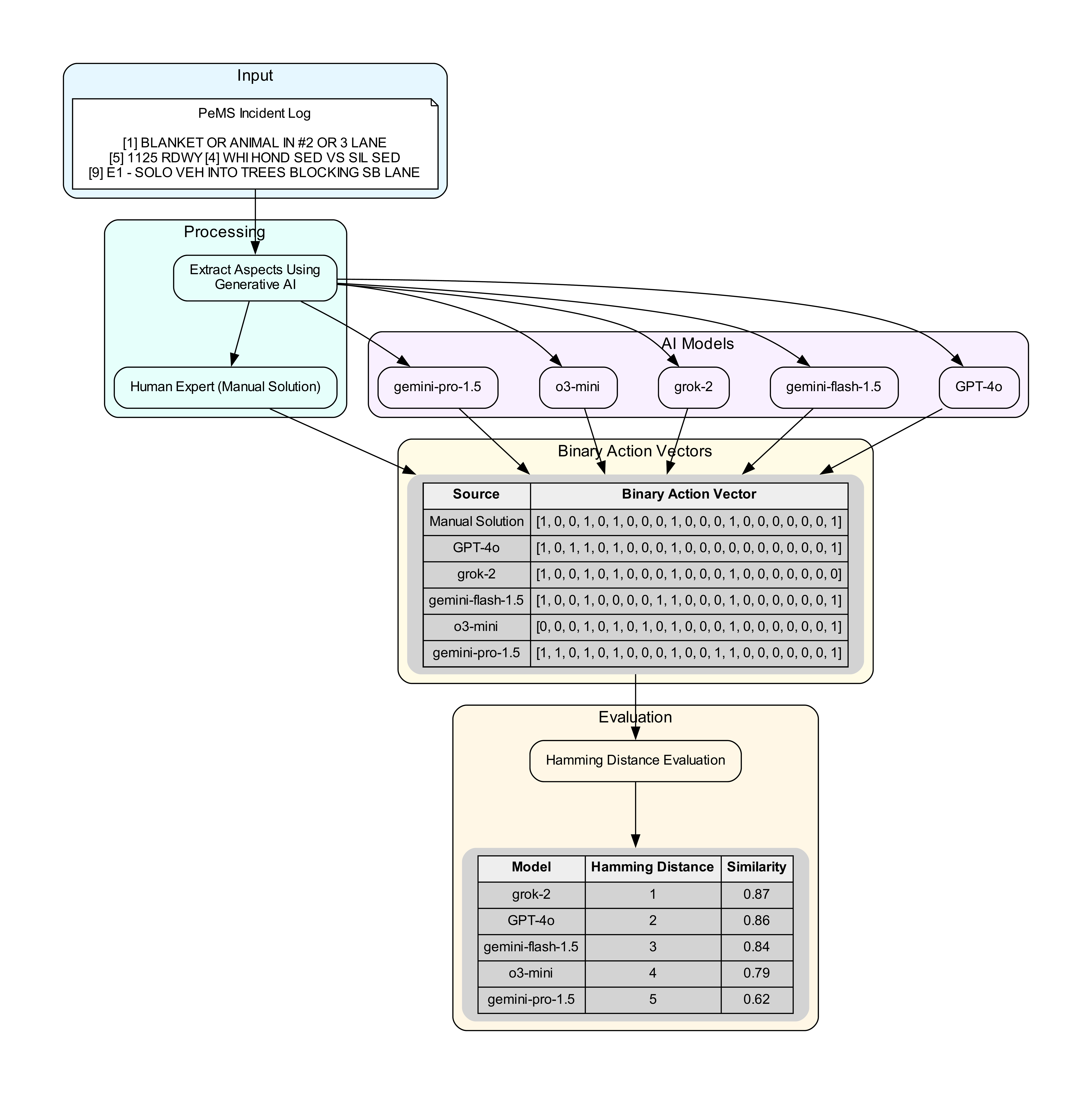}
    \caption{Evaluation pipeline for assessing AI model performance in traffic incident management action recommendation. The pipeline shows how PeMS incident logs flow through aspect extraction to binary action vectors for both manual solutions and various AI models. Hamming distance is used to calculate similarity between each model's recommendations and the expert solution.}
    \label{fig:pipeline}
\end{figure}

To evaluate how different AI models perform in recommending traffic incident management actions, we developed an evaluation pipeline illustrated in Figure~\ref{fig:pipeline}. The pipeline begins with raw PeMS (Performance Measurement System) incident logs that contain unstructured text descriptions of traffic incidents. These descriptions include information about accidents, hazards, and emergency response details. Using generative AI, we extract meaningful aspects from these incident logs to create a structured representation of each incident together with a reference set of appropriate actions for each incident, creating binary action vectors where each element represents whether a specific action (e.g., "Request Towing Services" or "Deploy Cones/Flares") should be taken. In parallel, we prompt several state-of-the-art generative AI models including GPT-4o, grok-2, gemini-flash-1.5, o3-mini, and gemini-pro-1.5 to produce their own binary action vectors for the same incidents. We then evaluate each model's performance by calculating the Hamming distance between its predictions and the manual (expert solutions) reference vectors. This distance measures how many action recommendations differ between the model and human experts, with lower values indicating better alignment with expert judgment. The evaluation results reveal varying levels of agreement between different models and the manual solution.

Our analysis using this pipeline allowed us to identify significant patterns in model behavior. The evaluation results also highlight interesting discrepancies in how different models interpret certain actions, with some models showing distinctive patterns of over-recommendation or under-recommendation for specific action categories.

\subsection{Traffic Incident Reports}
\label{sec:pems-case-study}

Caltrans' Performance Measurement System (PeMS) is a platform that collects and analyzes real-time and historical traffic data across California's freeway network. Among its various data components, PeMS includes detailed records of traffic incidents, which are critical for understanding and managing roadway conditions.

PeMS integrates traffic incident information from multiple sources, notably the California Highway Patrol's (CHP) Computer-Aided Dispatch (CAD) system. This integration provides a rich dataset that describes various aspects of traffic incidents, including:

\begin{itemize}
    \item \textbf{Incident Type:} Classification of the incident (e.g., collision, stalled vehicle, debris on roadway).
    \item \textbf{Location:} Precise location details, often including freeway segment and direction.
    \item \textbf{Time and Date:} Timestamp of when the incident was reported and cleared.
    \item \textbf{Severity:} Information on the seriousness of the incident, such as the number of lanes affected or if there were injuries.
    \item \textbf{Response Actions:} Details of the response, including agencies involved and measures taken to resolve the incident.
\end{itemize}

The benchmark dataset is sourced from the Caltrans Performance Measurement System (PeMS), comprises of a collection of 200 traffic incident reports that capture the evolution of roadway emergencies. Each record in the dataset typically includes an incident identifier, precise timestamps, location or sensor identifiers, and a mixture of structured status updates (e.g., “Unit Assigned,” “Unit Cleared,” “Unit Enroute”) alongside unstructured narrative text. For example, sample records from the dataset include entries such as:

\begin{itemize}
    \item \texttt{21441382,85811507,02/01/2023 16:11:00,Unit Assigned}
    \item \texttt{21441381,85811520,02/01/2023 16:11:00,Unit Cleared}
    \item \texttt{21441314,85811526,02/01/2023 16:11:00,[5] [Notification] [FSP]-REQ BREAK SB 215 JNO INLAND \\
    CENTER TO GET VEH TO RS [Shared]}
    \item \texttt{21441239,85811528,02/01/2023 16:11:00,[28] [FSP] has closed their incident [230201INFSP00241]}
\end{itemize}

These examples illustrate the dual nature of the logs: while the structured elements provide clear operational status (e.g., dispatch, en route, on scene, clearance), the unstructured components often embed critical details regarding incident specifics (such as vehicle conditions, communication updates, or equipment requirements). This blend of data types makes the dataset particularly challenging to analyze using traditional rule-based methods alone.

\subsection{Extracting Accident Descriptions and Actions from PeMS Logs}
\label{sec:accident-extraction}

The core of our methodology involves a two-step strategy. First, we use targeted parsing to isolate the free-form textual accident characteristics of the accident from the broader log entries. As illustrated by the provided examples, these narratives often contain critical details regarding the nature of the incident (e.g., vehicle types, objects involved, location specifics, initial impact assessment). The extraction process is designed to identify and retain these descriptive text segments.

Second, we address the extraction of operational actions. This is achieved through a combination of direct log analysis and inferential reasoning (using LLM capabilities) based on the textual accident narratives. Certain actions are explicitly stated within the structured or semi-structured components of the log (e.g., "Unit Enroute," "Unit Assigned," "Unit Cleared"). However, a significant aspect of our methodology involves inferring implied actions from the free-form text. To facilitate this, we define a predefined taxonomy of relevant accident response actions (e.g., "Initiate Lane Closures," "Request Towing Services," "Dispatch Police or EMS," "Request Animal Control"). This taxonomy serves as the basis for a targeted information extraction process.

The foundation of our action extraction process lies in a manualy derived taxonomy of relevant operational actions (see Table \ref{tab:predefined_actions}). It was developed through an initial, in-depth observation and analysis of approximately 1,000 randomly selected PeMS accident reports.

\begin{table}[htbp]
    \centering
    \caption{Predefined Taxonomy of Accident Response Actions}
    \label{tab:predefined_actions}
    \begin{tabular}{@{}p{7cm} p{7cm}@{}}
        \toprule
        Action & Action \\
        \midrule
        Activate VMS & Request Hazmat Team \\
        Implement Detours & Request Towing Services \\
        Initiate Lane Closures & Dispatch Police or EMS \\
        Remove Debris Hazards & Request Ambulance or Medical \\
        Deploy Cones/Flares & Request Fire Department or Rescue \\
        Request Occupant Extrication & Request Occupant Transport \\
        Implement Traffic Breaks & Notify TMC \\
        Issue SigAlert & Notify Local Police Department \\
        Notify Utility Companies & Request Animal Control \\
        Request Heavy or Specialized Tow & Request Lab Tech (Phlebotomy) \\
        Evacuate or Secure & \\
        \bottomrule
    \end{tabular}
\end{table}

To automate the identification of these actions, we utilize a prompt-based approach, specifically designed for large language models (LLMs). The provided prompt exemplifies this strategy, where an accident log is presented to the LLM along with the predefined list of actions. The LLM is instructed to generate a binary vector (represented as a Python dictionary) indicating the presence (1) or absence (0) of each action within the given log. This approach relies on the contextual understanding capabilities of LLMs to identify both explicitly mentioned actions (e.g., a direct request for a tow truck) and implicitly suggested actions (e.g., "BLOCKING SB LANE" implying "Initiate Lane Closures").

The prompt design uses clear instructions and provides a structured format (Python dictionary example) with a predefined list of actions. This direct approach guides the LLM to output binary flags (0/1) indicating the presence of each action in the accident log, ensuring a consistent and machine-readable result. The prompt explicitly restricts the LLM to the provided action list and output format.

In summary, our methodology for extracting accident descriptions and actions from PeMS logs comprises: (i) targeted parsing of free-form narrative text to capture accident characteristics, and (ii) a hybrid approach to action identification, combining the extraction of explicitly stated actions using a predefined taxonomy and LLM-based prompting. This forms a foundational dataset for subsequent LLM model evaluation on the task of incident response generation.

Although the PeMS logs contain a wide range of operational information (such as call times, unit dispatch, and clearance timestamps), this project focuses specifically on extracting the textual content that describes the accident’s details. The PeMS dataset, provided by Caltrans, offers an overview of traffic incidents by combining structured status updates with unstructured accident narratives. For this study, we randomly selected 200 accident records from the dataset (records from the February 2023). The following short excerpts illustrate how both the structured operational data and the free-form narrative are represented within a single record.

\begin{itemize}
    \item \textbf{Description 1:} \texttt{[1] BLANKET OR ANIMAL IN \#2 OR 3 LANE [5] 1125 RDWY [4] WHI HOND SED VS SIL SED Unit Enroute Unit Assigned Unit Enroute Unit Assigned Unit Cleared Unit Assigned Unit Enroute Unit At Scene [9] E1 - SOLO VEH INTO TREES BLOCKING SB LANE / 1185 EXP [10] [Rotation Request Comment] 1039 SEBASTOPOL TOW 707-823-1061 Unit Cleared [14] REQ ETA FOR 1185 Unit Cleared [19] [Notification] [CHP]-Problem changed from 1183-Trfc Collision-Unkn Inj to 1182-Trfc Collision-No Inj by CHP [Shared]}
    \item \textbf{Description 2:} \texttt{[2] ON OFR [1] VEH FACING SIDEWAYS [5] INV PTY ORIG XFR'D TO OPD // INV PTY ADV OFF OFR BUT HAD LANG BARRIER Unit Enroute Unit Assigned [7] 1039 1141 Unit Cleared [8] C96-177T  --97-- BLKNG LN 1/2 OF OFR Unit At Scene [10] [Notification] [CHP]-Problem changed from 1183-Trfc Collision-Unkn Inj to 1182-Trfc Collision-No Inj by CHP [9] 96-177T SOLO 1182 /NEG INJURY [11] 177T VEH MOVED AWAY FROM SCENE TO ONR / NEG INJURIES Unit Cleared Unit Assigned}
\end{itemize}

From these narratives, the system selectively extracts \emph{characteristics} of the accident, such as:

\begin{itemize}
    \item \textbf{Description 1 Characteristics:} ``BLANKET OR ANIMAL IN \#2 OR 3 LANE,'' ``1125 RDWY,'' ``WHI HOND SED VS SIL SED,'' ``E1 - SOLO VEH INTO TREES BLOCKING SB LANE / 1185 EXP''
    \item \textbf{Description 2 Characteristics:} ``[2] ON OFR,'' ``[1] VEH FACING SIDEWAYS,'' ``[5] INV PTY ORIG XFR'D TO OPD // INV PTY ADV OFF OFR BUT HAD LANG BARRIER,'' ``[7] 1039 1141,'' ``[8] C96-177T --97-- BLKNG LN 1/2 OF OFR,'' ``[9] 96-177T SOLO 1182 /NEG INJURY,'' ``[11] 177T VEH MOVED AWAY FROM SCENE TO ONR / NEG INJURIES''
\end{itemize}

Beyond these textual descriptions, each record often indicates any \emph{actions} taken or recommended by the responding units. We collect these actions—both directly from the log (e.g., ``Unit Enroute,'' ``Request Towing Services'') and from manual reviews of the free-form text. Below are shortened examples of \emph{manual annotations} representing whether an action was judged relevant (1) or not relevant (0):

\begin{lstlisting}[style=mystyle, language=Python, caption={Accident Log Processing Prompt}]
manual_actions = {
    "Activate VMS": 0,
    "Implement Detours": 0,
    "Initiate Lane Closures": 1,  # Implied by blocking SB lane
    ...
    "Request Towing Services": 1,  # Tow requested in [10]
    "Dispatch Police or EMS": 1,   # Units enroute imply dispatch
    "Request Animal Control": 1,   # Animal in the road
    ...
}
\end{lstlisting}


These dictionary-like annotations record which actions (e.g., \emph{Initiate Lane Closures}, \emph{Request Towing Services}) are triggered under different circumstances. In other words, each PeMS log is processed to retain only the accident-related text and any identified or inferred actions. This curated information serves as the primary input for subsequent phases in our workflow.

Below is the prompt that used to extract the relevant actions from the accident report:

\begin{lstlisting}[style=mystyle, language=Python, caption={Accident Log Processing Prompt}]
r"""
Accident log:

{log}

Task:
------
Based on the provided accident log, generate a binary vector indicating whether each predefined accident action is observed in the log.

Predefined Actions:
-------------------
- "Activate VMS"
- "Implement Detours"
- "Initiate Lane Closures"
- "Remove Debris Hazards"
- "Deploy Cones/Flares"
- "Request Hazmat Team"
- "Request Towing Services"
- "Dispatch Police or EMS"
- "Request Ambulance or Medical"
- "Request Fire Department or Rescue"
- "Request Occupant Extrication"
- "Request Occupant Transport"
- "Implement Traffic Breaks"
- "Notify TMC"
- "Issue SigAlert"
- "Notify Local Police Department"
- "Notify Utility Companies"
- "Request Animal Control"
- "Request Heavy or Specialized Tow"
- "Request Lab Tech (Phlebotomy)"
- "Evacuate or Secure"

Instructions:
-------------
1. Parse the accident log to determine which actions are mentioned.
2. For each predefined action, assign a binary value: 1 if the action is present in the log; 0 otherwise.
3. Do not invent any new actions or modify the list provided.
4. Your final output must be a Python dictionary exactly in the format shown below, and nothing else.

Example Response:
-----------------
actions = {
    "Activate VMS": 0,
    "Implement Detours": 1,
    "Initiate Lane Closures": 1,
    "Remove Debris Hazards": 1,
    "Deploy Cones/Flares": 0,
    "Request Hazmat Team": 1,
    "Request Towing Services": 0,
    "Dispatch Police or EMS": 1,
    "Request Ambulance or Medical": 1,
    "Request Fire Department or Rescue": 1,
    "Request Occupant Extrication": 1,
    "Request Occupant Transport": 1,
    "Implement Traffic Breaks": 0,
    "Notify TMC": 0,
    "Issue SigAlert": 0,
    "Notify Local Police Department": 0,
    "Notify Utility Companies": 0,
    "Request Animal Control": 0,
    "Request Heavy or Specialized Tow": 0,
    "Request Lab Tech (Phlebotomy)": 0,
    "Evacuate or Secure": 1
}

Response (provide the Python dictionary only, following the example):
"""
\end{lstlisting}

Below is the prompt that used to extract the relevant description tags from the accident report:

\begin{lstlisting}[style=mystyle, caption={Accident Characteristics Extraction Prompt}]
Accident log:

{log}

Given the accident log, copy only parts of text that contains only accident descriptions. 
Exclude any actions by traffic agency or response party (e.g. managing alerts).

Response (only comma-separated quotes from the original accident log):
\end{lstlisting}

Here we highlight the necessity to use more complex (including reasoning) models. The provided prompt instructs the model to extract only the portions of the accident log that contain accident descriptions, while specifically omitting any text related to actions taken by traffic agencies or response parties. This task is not merely about pattern matching; it requires the model to perform nuanced reasoning and natural language understanding. This level of text parsing and contextual interpretation showcases the necessity for more advanced, complex models—ones that possess strong reasoning capabilities—to effectively process and extract the elements within real-world accident reports.

\begin{lstlisting}[frame=single, breaklines=true, caption={Prompt for Generating an example Incident Response Plan for Incident ID 21455531 (PeMS data set)}]
[1] BLANKET OR ANIMAL IN #2 OR 3 LANE [5] 1125 RDWY [4] WHI HOND SED VS SIL SED Unit Enroute Unit Assigned Unit Enroute Unit Assigned Unit Cleared Unit Assigned Unit Enroute Unit At Scene [9] E1 - SOLO VEH INTO TREES  BLOCKING SB LANE / 1185 EXP [10] [Rotation Request Comment] 1039 SEBASTOPOL TOW 707-823-1061 Unit Cleared [14] REQ ETA FOR 1185 Unit Cleared [19] [Notification] [CHP]-Problem changed from 1183-Trfc Collision-Unkn Inj to 1182-Trfc Collision-No Inj by CHP [Shared]

Based on the provided accident log, generate an incident response plan.
\end{lstlisting}

\begin{table}[h!]
\scriptsize
    \centering
    \begin{tabular}{|c|c|}
        \hline
        \textbf{ChatGPT 4o Response} & \textbf{Gemini Flash 1.5 Response} \\
        \hline
        \begin{minipage}[t]{0.45\textwidth}
            \textbf{Incident Response Plan: Traffic Collision - No Injuries} \\[5pt]
            \textbf{Incident Summary} \\
            \textbf{Type of Incident:} Traffic Collision – No Injuries (CHP Code 1182) \\
            \textbf{Location:} Southbound lane (exact location unspecified) \\
            \textbf{Involved Vehicles:} \\
            - White Honda Sedan \\
            - Silver Sedan \\
            \textbf{Additional Hazards:} \\
            - Vehicle into trees, blocking the southbound lane \\
            - Possible debris (blanket or animal) in \#2 or \#3 lane \\

            \textbf{Response Actions} \\[5pt]
            \textbf{Initial Assessment \& Dispatch} \\
            - Confirm the exact location using dispatch logs and GPS coordinates. \\
            - Assign appropriate units: \\
              - CHP Officers for traffic control and scene assessment. \\
              - Fire/EMS (if necessary) for potential vehicle hazards. \\
              - Tow Service (1185 Request) for vehicle removal. \\

            \textbf{Traffic Control \& Safety Measures} \\
            - Deploy traffic cones and flares to divert traffic around the incident. \\
            - CHP officers direct traffic and assess the extent of blockage. \\
            - Identify and remove any hazardous debris (blanket or animal) from \#2 or \#3 lane. \\

            \textbf{Vehicle Removal \& Road Clearance} \\
            - Tow service (Sebastopol Tow, 707-823-1061) en route for vehicle recovery. \\
            - Request estimated time of arrival (ETA) for tow. \\
            - Officers confirm no fluid leaks or additional road hazards before reopening lanes. \\

            \textbf{Status Update \& Finalization} \\
            - Notify all responding units that the incident has been downgraded to No Injuries (1182). \\
            - Ensure all documentation is updated to reflect the correct incident classification. \\
            - Clear all units from the scene once road conditions are deemed safe. \\

            \textbf{Follow-Up Actions} \\
            - \textbf{CHP Documentation:} Ensure detailed incident report completion, including driver statements and possible contributing factors. \\
            - \textbf{Tow Coordination:} Confirm vehicle recovery status with Sebastopol Tow. \\
            - \textbf{Roadway Inspection:} Confirm no remaining debris or damage that could affect future traffic flow. \\

            This plan ensures a swift, organized response to the incident while minimizing traffic disruption and ensuring public safety.
        \end{minipage}
        &
        \begin{minipage}[t]{0.45\textwidth}
            \textbf{Incident Response Plan: Single Vehicle Accident at 1185 Expressway} \\[5pt]
            \textbf{Incident:} Single vehicle collision into trees, blocking the southbound lane at approximately 1185 Expressway (1125 Roadway is likely a nearby reference point). \\

            \textbf{Initial Assessment:} \\
            - \textbf{Severity:} Moderate (single vehicle, trees involved, lane blockage). Further assessment needed regarding injuries (initial report suggested unknown, later revised to no injuries). \\
            - \textbf{Location:} 1185 Expressway, Southbound lanes blocked. \\
            - \textbf{Resources:} E1 (Engine 1) is en route and assigned. Tow truck (Sebastopol Tow, 707-823-1061) has been contacted. CHP is involved and has updated the incident classification to "no injury." \\

            \textbf{Response Plan:} \\

            \textbf{Phase 1: Immediate Response (Ongoing)} \\
            - \textbf{Scene Security \& Safety:} E1 will secure the scene, directing traffic around the blockage and ensuring the safety of first responders and the public. \\
            - \textbf{Casualty Care:} While the CHP has revised the incident to "no injury", E1 should conduct a thorough assessment of the driver and any passengers for injuries. Provide any necessary first aid. \\
            - \textbf{Initial Assessment:} E1 will assess the extent of damage to the vehicle and the environment (trees). Determine if any hazardous materials are involved. \\

            \textbf{Phase 2: Incident Mitigation} \\
            - \textbf{Traffic Management:} Maintain traffic control until the roadway is cleared. This may involve coordinating with CHP. \\
            - \textbf{Vehicle Removal:} Coordinate with Sebastopol Tow (707-823-1061) to remove the vehicle. Confirm ETA and facilitate efficient removal. Request an update on ETA if necessary. \\
            - \textbf{Debris Removal:} Clear any debris from the roadway, including broken glass, vehicle parts, and branches. \\

            \textbf{Phase 3: Post-Incident} \\
            - \textbf{Scene Clearance:} Ensure the area is completely clear and safe for normal traffic flow. \\
            - \textbf{Report Completion:} E1 will complete a thorough incident report, including details of damage, injuries (or lack thereof), and actions taken. Include details about the tow truck's involvement. \\
            - \textbf{Follow-up:} If necessary, follow up with the CHP regarding their investigation. \\

            \textbf{Communication:} \\
            - Maintain clear and concise communication between E1, the tow truck company, and CHP throughout the incident. Use radio communication and phone calls as needed. \\
            - Provide regular updates to dispatch on progress and any changes to the situation. \\

            \textbf{Contingency Planning:} \\
            - If injuries are discovered despite initial reports, immediately request additional medical resources (ambulance). \\
            - If hazardous materials are involved, implement appropriate protocols and request specialized HAZMAT teams. \\
            - If the incident significantly impacts traffic flow, consider notifying local traffic management authorities for broader traffic control strategies. \\

            This plan is based on the limited information provided. The actual response may need adjustment based on conditions at the scene.
        \end{minipage} \\
        \hline
    \end{tabular}
    \caption{Comparison of Incident Response Plans from ChatGPT 4o and Gemini Flash 1.5}
    \label{tab:incident_response_comparison}
\end{table}

Table~\ref{tab:incident_response_comparison} presents a side-by-side comparison of raw incident response plans generated by two AI models: ChatGPT 4o and Gemini Flash 1.5. The left column details ChatGPT 4o’s narrative response, which includes an incident summary, a breakdown of response actions, and follow-up procedures. In contrast, the right column displays Gemini Flash 1.5’s plan, which is organized into distinct phases—covering immediate response, incident mitigation, and post-incident activities - with clearly delineated steps and resource coordination. This juxtaposition highlights the inherent differences in the models’ to structuring its responses and prioritizing various aspects of incident management.

\subsection{Incident response plan comparison methodology}

During the incident management, decision-makers are frequently confronted with the need to evaluate and select from a multitude of response actions under conditions of uncertainty and time constraints. The commonly used method is Analytical Hierarchy Process (AHP) which serves as a robust multi-criteria decision-making (MCDM) tool that assists in organizing and analyzing these complex decisions. By breaking down the decision-making problem into a hierarchy of more easily comprehended sub-problems, AHP enables decision-makers to systematically evaluate the relative importance of various criteria and actions. 
While the Analytical Hierarchy Process (AHP) provides a structured approach for multi-criteria decision-making (MCDM), TOPSIS \cite{LAI1994486,zavadskas2016development} is another robust MCDM technique that ranks alternatives using multiple criteria based on their relative closeness to an ideal solution. Unlike AHP, which involves pairwise comparisons and eigenvector calculations, TOPSIS evaluates alternatives based on their geometric distance to a positive ideal solution (PIS) and a negative ideal solution (NIS).

In this study, we adopt a simplified methodology based on non-weighted binary vectors to represent incident response plans and evaluate them using Hamming distance. This method disregards any associated weights or priority levels and focuses solely on the presence or absence of actions within a plan. Each incident response plan is represented as a binary vector, where each element corresponds to a specific action, such as "Deploy Incident Response Vehicle," "Temporary Lane Closure," or "Notify Social Media." A value of 1 indicates the inclusion of an action in the plan, while a value of 0 indicates its exclusion. Similarity or difference between plans is then measured using the Hamming distance, which counts the number of differing elements between two binary vectors. The Hamming distance serves as a straightforward metric to compare the structure of response plans and assess how one plan diverges from another.

The key advantage of using non-weighted binary vectors and Hamming distance lies in its simplicity and objectivity. Since this methodology does not rely on action weights or subjective criteria, it avoids the complexities associated with assigning importance to individual actions. This is particularly useful when domain knowledge about the relative impact or resource requirements of actions is limited or unavailable.

By focusing exclusively on the structure of action sets, this approach enables the straightforward identification of similarities and differences between response plans.

Our method to compare these binary action vectors is by calculating their Hamming distance. The Hamming distance between two binary vectors of equal length is the number of positions at which the corresponding symbols are different.

Mathematically, the Hamming distance \( d_H \) between two binary vectors \( \mathbf{u} = (u_1, u_2, \ldots, u_n) \) and \( \mathbf{v} = (v_1, v_2, \ldots, v_n) \) is defined as:

\[
d_H(\mathbf{u}, \mathbf{v}) = \sum_{i=1}^{n} |u_i - v_i|
\]

In the context of comparing incident response plans (represented as binary action vectors), the Hamming distance provides a measure of how many actions differ between the two plans and offers a straightforward way to quantify the dissimilarity between different proposed incident response plans based solely on the presence or absence of specific actions. A smaller Hamming distance indicates a greater similarity in the set of actions included in the two plans. For example, comparing Plan A and Plan B from \autoref{lst:example_plans}:

\begin{lstlisting}[style=mystyle, language=, caption={Example Incident Response Plans as Binary Vectors}, label={lst:example_plans}]
Plan A: [1, 0, 1, 1, 0, 1, 0, 1, 0, 1]
Plan B: [1, 1, 0, 0, 1, 0, 1, 0, 1, 0]
\end{lstlisting}

The differences occur at the 2nd, 3rd, 4th, 5th, 7th, 9th, and 10th positions. Therefore, the Hamming distance between Plan A and Plan B is 7.

\subsection{Late fusion of model responses}

The use of binary action decomposition enables the effective application of ensemble techniques to improve the accuracy of response plans. By converting model outputs into binary action vectors, we can treat each model's response as a feature vector. This allows for the application of ensembling methods, such as late fusion, where multiple models’ outputs are combined to produce a more optimal averaged solution \cite{akilan2018effect,boulahia2021early}.

Using multiple executions of the prompt we may obtain multiple binary vectors (solutions): \( \mathbf{v_1}, \mathbf{v_2}, \dots, \mathbf{v_m} \), where \( m \) is the number of responses, and each vector is of length \( n \):

\[
\mathbf{v_1} = [v_{11}, v_{12}, \dots, v_{1n}]
\]
\[
\mathbf{v_2} = [v_{21}, v_{22}, \dots, v_{2n}]
\]
\[
\vdots
\]
\[
\mathbf{v_m} = [v_{m1}, v_{m2}, \dots, v_{mn}]
\]

The element-wise average of these \( m \) vectors is given by:

\[
\mathbf{avg} = \frac{1}{m} \left( \mathbf{v_1} + \mathbf{v_2} + \dots + \mathbf{v_m} \right)
\]

This can be written as:

\[
\mathbf{avg} = \left[ \frac{v_{11} + v_{21} + \dots + v_{m1}}{m}, \frac{v_{12} + v_{22} + \dots + v_{m2}}{m}, \dots, \frac{v_{1n} + v_{2n} + \dots + v_{mn}}{m} \right]
\]

Next, apply a rounding function, \( \text{round}(x) \), to each element of the averaged vector to obtain the final binary vector \( \mathbf{b} \):

\[
\mathbf{b} = \text{round}\left( \mathbf{avg} \right)
\]

This results in:

\[
\mathbf{b} = \left[ \text{round}\left( \frac{v_{11} + v_{21} + \dots + v_{m1}}{m} \right), \text{round}\left( \frac{v_{12} + v_{22} + \dots + v_{m2}}{m} \right), \dots, \text{round}\left( \frac{v_{1n} + v_{2n} + \dots + v_{mn}}{m} \right) \right]
\]

The resulting vector \( \mathbf{b} \) is the final binary decision vector derived by averaging and rounding the outputs from the \( m \) binary vectors.

For example, let's consider binary responses from three models for a single incident:

\[
\mathbf{r_1} = [1, 0, 1, 1, 0]
\]
\[
\mathbf{r_2} = [1, 1, 0, 1, 1]
\]
\[
\mathbf{r_3} = [0, 1, 1, 0, 0]
\]

Element-wise summation:
\[
\mathbf{sum} = \mathbf{r_1} + \mathbf{r_2} + \mathbf{r_3} = [1+1+0, 0+1+1, 1+0+1, 1+1+0, 0+1+0]
\]
\[
\mathbf{sum} = [2, 2, 2, 2, 1]
\]

Step 2: Compute the average:
\[
\mathbf{avg} = \frac{1}{3} \mathbf{sum} = \left[ \frac{2}{3}, \frac{2}{3}, \frac{2}{3}, \frac{2}{3}, \frac{1}{3} \right]
\]

Step 3: Rounding to final binary response:
\[
\mathbf{final} = \text{round}(\mathbf{avg}) = [1, 1, 1, 1, 0]
\]

The final binary response is \( \mathbf{final} = [1, 1, 1, 1, 0] \), representing the averaged decision from all models.

This approach has the potential to reduce minor reasoning errors and enhance the stability of the final generated plan. Late fusion is also critical for minimizing randomness in the generated responses and ensuring the reproducibility of results. As we propose a method for quantifying solutions based on their proximity to a manual solution, the impact of different numbers of plan generations on overall model performance can also be assessed.



In the Table \ref{tab:modresp}, each row represents a specific action (for example, \emph{Activate VMS} or \emph{Request Animal Control}), while each column corresponds to a different model — ranging from the manual evaluation to various AI models (GPT-4o, 4o-mini, o1-mini, Gemini-Flash-2, Gemini-Flash-1.5, Gemini-Pro-1.5). The binary values (0 or 1) indicate whether an action is not recommended (0) or recommended (1). For instance, in the \emph{Activate VMS} row, only 4o-mini, o1-mini, and Gemini-Pro-1.5 recommend activating VMS (indicated by a 1), while the manual evaluation and other models do not.

\begin{table}[htbp]
\small  
\begin{tabular}{lccccccc}
\hline
\textbf{Action Key} & \textbf{Manual} & \textbf{GPT-4o} & \textbf{4o-mini} & \textbf{o1-mini} & \textbf{G-Flash-2} & \textbf{G-Flash-1.5} & \textbf{G-Pro-1.5} \\ \hline
Activate VMS & 0 & 0 & 1 & 1 & 0 & 0 & 1 \\
Deploy Cones/Flares & 0 & 0 & 0 & 0 & 0 & 0 & 1 \\
Dispatch Police or EMS & 1 & 0 & 1 & 1 & 1 & 1 & 1 \\
Evacuate or Secure & 0 & 0 & 0 & 1 & 0 & 0 & 1 \\
Implement Detours & 0 & 0 & 0 & 1 & 0 & 0 & 1 \\
Implement Traffic Breaks & 0 & 0 & 0 & 0 & 0 & 0 & 1 \\
Initiate Lane Closures & 1 & 1 & 1 & 0 & 1 & 1 & 1 \\
Issue SigAlert & 0 & 0 & 0 & 0 & 1 & 0 & 1 \\
Notify Local Police Department & 0 & 0 & 1 & 0 & 1 & 0 & 1 \\
Notify TMC & 0 & 0 & 0 & 0 & 1 & 0 & 1 \\
Notify Utility Companies & 0 & 0 & 0 & 0 & 0 & 0 & 0 \\
Remove Debris Hazards & 0 & 0 & 1 & 1 & 1 & 1 & 1 \\
Request Ambulance or Medical & 0 & 0 & 0 & 0 & 1 & 0 & 1 \\
Request Animal Control & 1 & 1 & 1 & 1 & 1 & 1 & 1 \\
Request Fire Department or Rescue & 0 & 0 & 0 & 1 & 1 & 0 & 1 \\
Request Hazmat Team & 0 & 0 & 0 & 0 & 0 & 0 & 0 \\
Request Heavy or Specialized Tow & 0 & 0 & 0 & 1 & 0 & 0 & 0 \\
Request Lab Tech (Phlebotomy) & 0 & 0 & 0 & 0 & 0 & 0 & 0 \\
Request Occupant Extrication & 0 & 0 & 0 & 1 & 1 & 1 & 1 \\
Request Occupant Transport & 0 & 0 & 0 & 0 & 1 & 0 & 1 \\
Request Towing Services & 1 & 1 & 1 & 1 & 1 & 0 & 1 \\
\hline
\end{tabular}
\caption{Binary action values for ID=21455531}
\label{tab:modresp}
\end{table}

\section{Discussion}

We evaluate how closely different models—including GPT-4o, Grok 2, and Gemini 1.5 Pro—align with expert manual solutions when formulating response plans. Our methodology uses multiple complementary metrics: Hamming distance to measure quantitative deviation from manual responses, TOPSIS-based weighted differences to account for action prioritization, and granular analysis of both unnecessary and missed actions. These metrics collectively provide an assessment framework that captures both accuracy and prioritization.

\paragraph{Table \ref{tab:hamming_distances}: Average Hamming Distances Compared to Manual Solution}  
This table reports the mean Hamming distance between each model's response plan and the Manual Solution. A lower Hamming distance indicates that the model’s response is closer to the manual response. In this evaluation, Grok 2 (2.96) and GPT-4o (2.98) achieve the smallest differences, suggesting they closely mirror the manual solution. In contrast, Gemini 1.5 Pro exhibits the highest average Hamming distance (8.39), indicating that its responses deviate substantially from the optimal plan.


\paragraph{Table \ref{tab:unnecessary_actions}: Unnecessary Actions (Model has 1, Manual Solution has 0)}  
This table shows the number of superfluous actions included by each model that are absent from the Manual Solution. A model with a lower count is more precise in its selection. In this metric, GPT-4o (225) and Grok 2 (237) perform well by keeping unnecessary actions to a minimum. In stark contrast, Gemini 1.5 Pro records a very high count of 1547 unnecessary actions, reflecting a tendency to over-trigger responses.

\paragraph{Table \ref{tab:missed_actions}: Missed Actions (Model has 0, Manual Solution has 1)}  
This table captures the instances where the model omitted actions that are present in the Manual Solution. While fewer missed actions are generally positive, this metric should be interpreted alongside unnecessary actions. Notably, Gemini 1.5 Pro misses the fewest actions (132), which might seem advantageous at first glance. However, when combined with its excessive unnecessary actions, it indicates that Gemini 1.5 Pro often over-compensates—triggering many unneeded responses to ensure it does not miss an essential one. Other models such as GPT-4o and Grok 2 have moderate misses (372 and 356 respectively) but achieve a better overall balance between precision and completeness.

The analysis of both unnecessary and missed actions (Tables~\ref{tab:unnecessary_actions} and \ref{tab:missed_actions}) reveals complementary aspects of model performance. Although Gemini 1.5 Pro benefits from a low count of missed actions (132), its extremely high count of unnecessary actions (1547) suggests that it over-responds as a strategy to avoid omissions. GPT-4o, on the other hand, manages to keep both unnecessary actions (225) and missed actions (372) relatively balanced, which indicates a finely tuned response strategy. Grok 2 also demonstrates a close alignment in terms of minimal divergence, with a strong overall performance across the evaluated metrics.

In summary, while models such as GPT-4o and Grok 2 consistently produce response plans that are accurate, Gemini 1.5 Pro tends to over-trigger actions, compromising its overall effectiveness despite minimizing missed critical actions. These results highlight the importance of considering multiple performance dimensions when assessing the quality and alignment of automated incident response plans with human-generated solutions.

\begin{table}[ht]
\centering
\footnotesize
\begin{subtable}[t]{0.48\linewidth}
  \centering
  \caption{Hamming Distances}
  \label{tab:hamming_distances}
  \begin{tabular}{lr}
    \toprule
    Model & Avg. Hamming \\
    \midrule
    Gemini 1.5 Flash & 3.52 \\
    GPT-4o          & 2.98 \\
    GPT-4o Mini     & 4.99 \\
    GPT-o1 Mini     & 6.68 \\
    Gemini 1.5 Pro  & 8.39 \\
    Grok 2          & 2.96 \\
    \bottomrule
  \end{tabular}
\end{subtable}%
\hfill

\vspace{1ex}

\begin{subtable}[t]{0.48\linewidth}
  \centering
  \caption{Unnecessary Actions}
  \label{tab:unnecessary_actions}
  \begin{tabular}{lr}
    \toprule
    Model & Count \\
    \midrule
    Gemini 1.5 Flash & 403 \\
    GPT-4o          & 225 \\
    GPT-4o Mini     & 745 \\
    GPT-o1 Mini     & 1180 \\
    Gemini 1.5 Pro  & 1547 \\
    Grok 2          & 237 \\
    \bottomrule
  \end{tabular}
\end{subtable}%
\hfill
\begin{subtable}[t]{0.48\linewidth}
  \centering
  \caption{Missed Actions}
  \label{tab:missed_actions}
  \begin{tabular}{lr}
    \toprule
    Model & Count \\
    \midrule
    Gemini 1.5 Flash & 302 \\
    GPT-4o          & 372 \\
    GPT-4o Mini     & 253 \\
    GPT-o1 Mini     & 157 \\
    Gemini 1.5 Pro  & 132 \\
    Grok 2          & 356 \\
    \bottomrule
  \end{tabular}
\end{subtable}
\caption{Performance Comparison Based on Various Metrics}
\label{tab:all_metrics}
\end{table}

For detailed list of generated responses as well as considered accidents, please refer to the code: \url{https://github.com/Future-Mobility-Lab/TrafficIncidentGPT}.

\subsection{Model Ensemble Analysis Using Majority Voting}
In this study, we explore how combining multiple AI models through ensemble technique can improve the accuracy of traffic incident action recommendations. We evaluate all possible combinations of models using majority voting.

The majority voting ensemble approach works as follows:
\begin{enumerate}
    \item For each traffic incident, each individual model produces a binary action vector, where each element represents whether a specific action is recommended (1) or not (0).
    \item For a given ensemble of $n$ models, we collect their binary vectors for each incident.
    \item For each action, we calculate the average recommendation score by summing the binary values and dividing by the number of models in the ensemble.
    \item An action is included in the final ensemble recommendation if the average score exceeds 0.5 (indicating that more than half of the models recommend it).
    \item The resulting ensemble prediction is then compared to the manual solution (expert human judgment) using Hamming distance, which counts the number of positions where the vectors differ.
    \item Similarity is calculated as $1 - \frac{\text{Hamming distance}}{\text{total number of actions}}$.
\end{enumerate}

Instead of only evaluating sequential combinations, our approach examines \textit{all possible combinations} of models for each ensemble size, allowing us to identify the optimal groupings regardless of order.

\begin{figure}[htbp]
    \centering
    \includegraphics[width=0.8\textwidth]{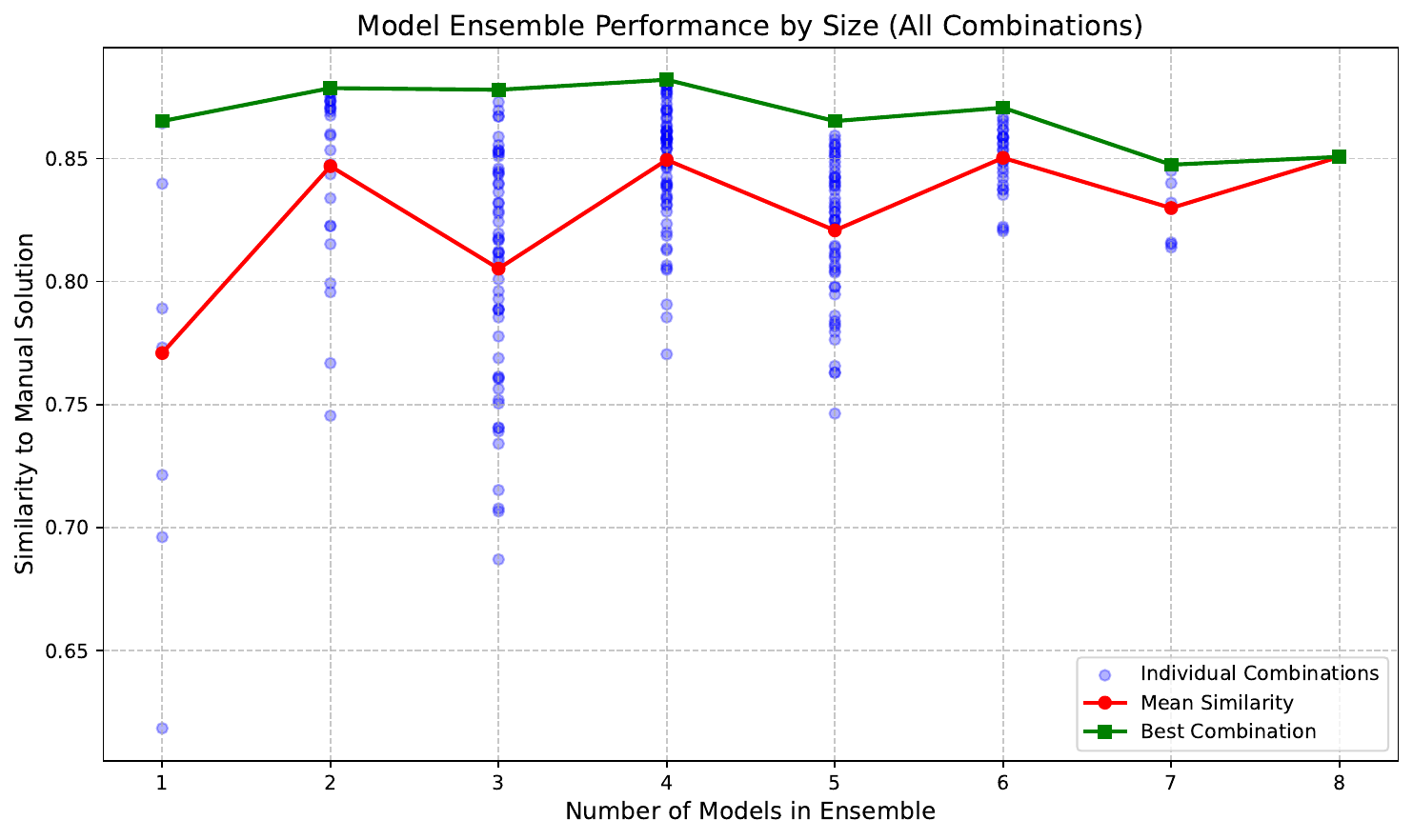}
    \caption{Model ensemble performance by size using majority voting. Each blue dot represents a specific model combination, with the red line showing average similarity across all combinations of that size. The green line highlights the best-performing combination for each size, with annotations indicating which specific models are included.}
    \label{fig:ensemble_performance}
\end{figure}

Figure~\ref{fig:ensemble_performance} presents the analysis of ensemble performance by size. Ideal similarity to manual solution is 1.0, but none of the models or their combinations reach it. The plot shows individual combinations as scatter points, with mean performance and best combination highlighted. The performance of individual models has the highest variance and we observe a significant boost in performance when combining any two models. However, the performance of some single models is suboptimal and including them in the ensemble may show the degradation of performance. There is a trend in reduction of variance and improvement in similarity to human expert solution when increasing the number of models in the ensemble. This indicates that majority voting is a highly relevant technique for critical recommendation tasks.

\begin{figure}[htbp]
    \centering
    \includegraphics[width=0.8\textwidth]{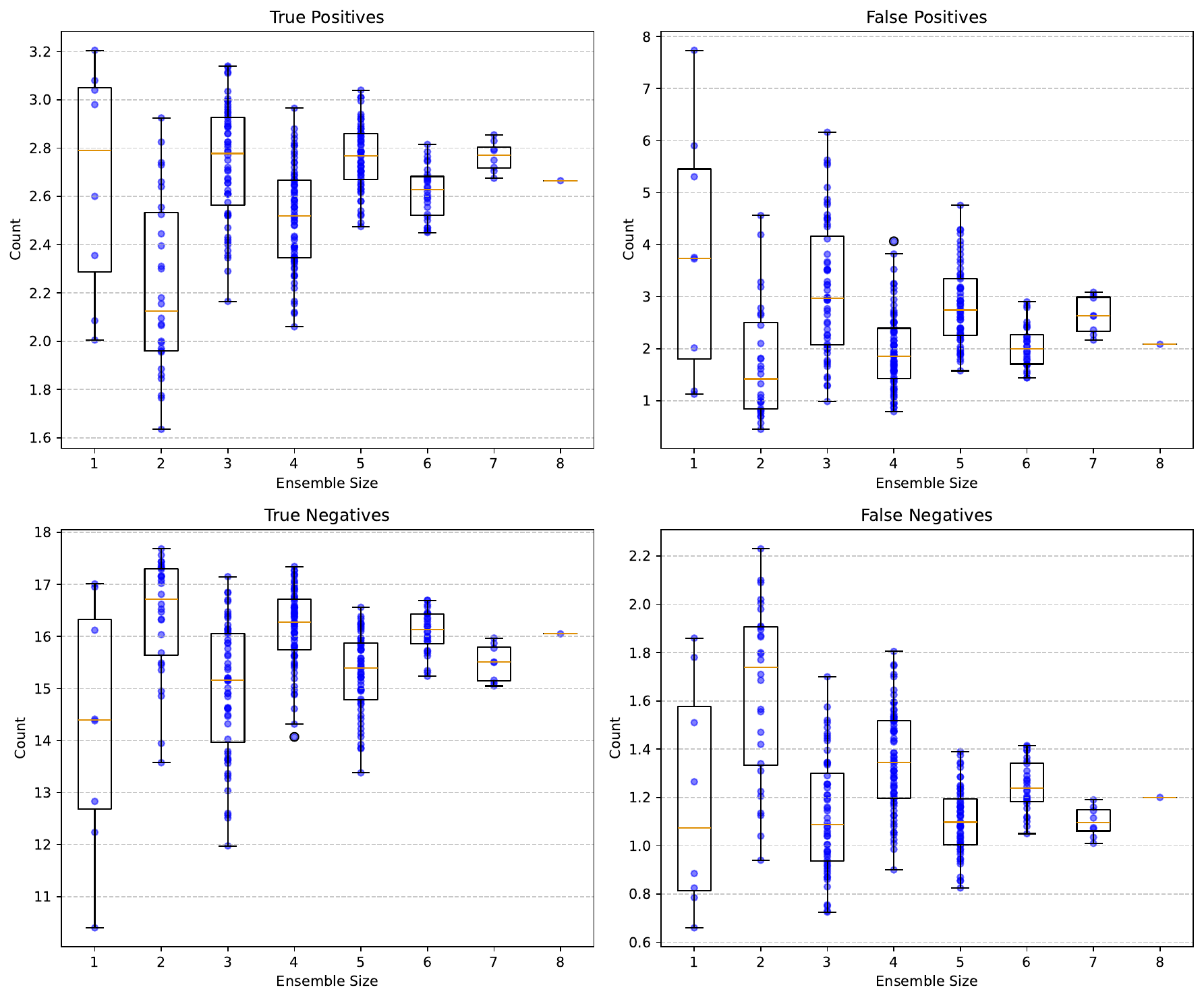}
    \caption{Box plots showing the distribution of performance metrics (similarity, precision, recall, and F1 score) across different ensemble sizes. The decreasing spread as ensemble size increases indicates more consistent performance with larger ensembles.}
    \label{fig:ensemble_boxplots}
\end{figure}

To better understand the statistical distribution of performance across different ensemble sizes, we analyze the distribution of similarity scores for each ensemble size using box plots of classification error rates (amount of true positives, true negatives, etc) across ensemble size, as shown in Figure~\ref{fig:ensemble_boxplots}. The total number of actions is 21. Analysis of the confusion matrix elements across different ensemble sizes reveals the following: the 4-model ensemble achieves the highest similarity score (0.882 - see Figure \ref{fig:ensemble_performance}) with an optimal balance of True Positives (2.06), low False Positives (0.79), good True Negatives (17.35), and moderate False Negatives (1.81), indicating it's the best overall configuration; even-numbered ensembles (2, 4, 6) consistently show better performance including lower False Positive rates than odd-numbered ones, suggesting the importance of choosing the voting approach.

Our analysis of majority voting ensembles demonstrates that combining multiple AI models significantly improves action recommendation quality for traffic incident management. The optimal approach appears to be using a selected ensemble of 3-4 models, which balances performance gains with computational efficiency.

The majority voting mechanism proves particularly effective at filtering out anomalous recommendations from individual models, resulting in prediction sets that more closely align with human expert judgment. This finding suggests that deployment of ensemble systems, rather than single models, may be the most reliable approach for safety-critical domains like traffic incident management.

Future work should explore more sophisticated response aggregation methods, such as weighted voting based on model confidence or domain-specific expertise, which might further improve performance in specific action categories.

\subsection{Model Agreement Analysis}
\begin{figure}[H]
    \centering
    \includegraphics[width=0.8\textwidth]{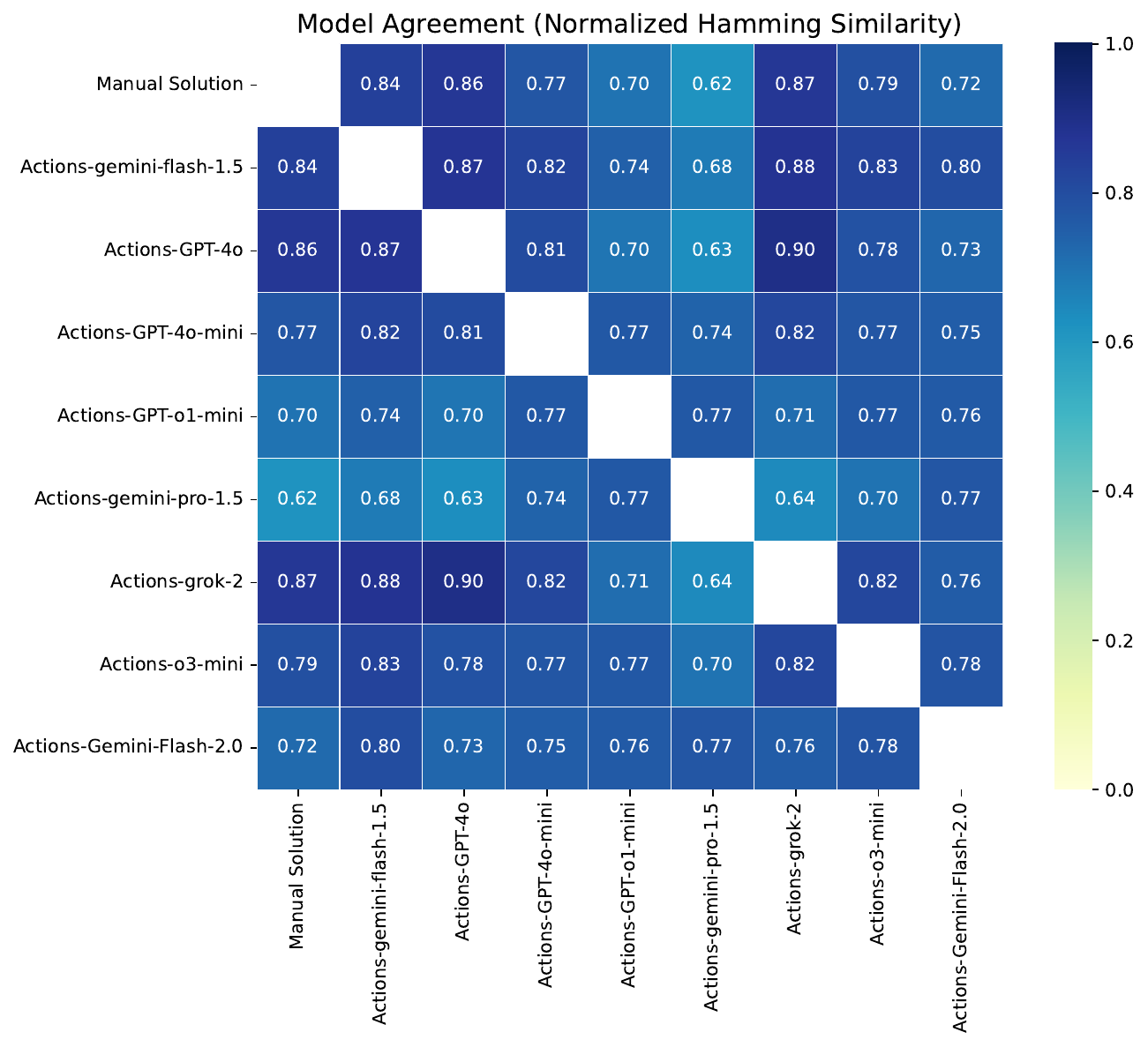}
    \caption{Model Agreement Matrix (Normalized Hamming Similarity)}
    \label{fig:similarity}
\end{figure}

Figure~\ref{fig:similarity} shows the normalized Hamming similarity between models, where higher values indicate greater agreement in action recommendations.  The analysis reveals that the models with the highest agreement with the Manual Solution are Actions-grok-2 (86.5\%), Actions-GPT-4o (86.4\%), and Actions-gemini-flash-1.5 (84.0\%), indicating robust performance among these systems. In contrast, Actions-gemini-pro-1.5 exhibits significantly lower alignment at only 61.8\%, suggesting limitations in its response generation. Moreover, clear model clusters emerge from the results; notably, GPT-4o and grok-2 form a high-similarity cluster, achieving an agreement rate of 89.9\%, which underscores the consistency in their decision-making processes.

\begin{figure}[H]
    \centering
    \includegraphics[width=\textwidth]{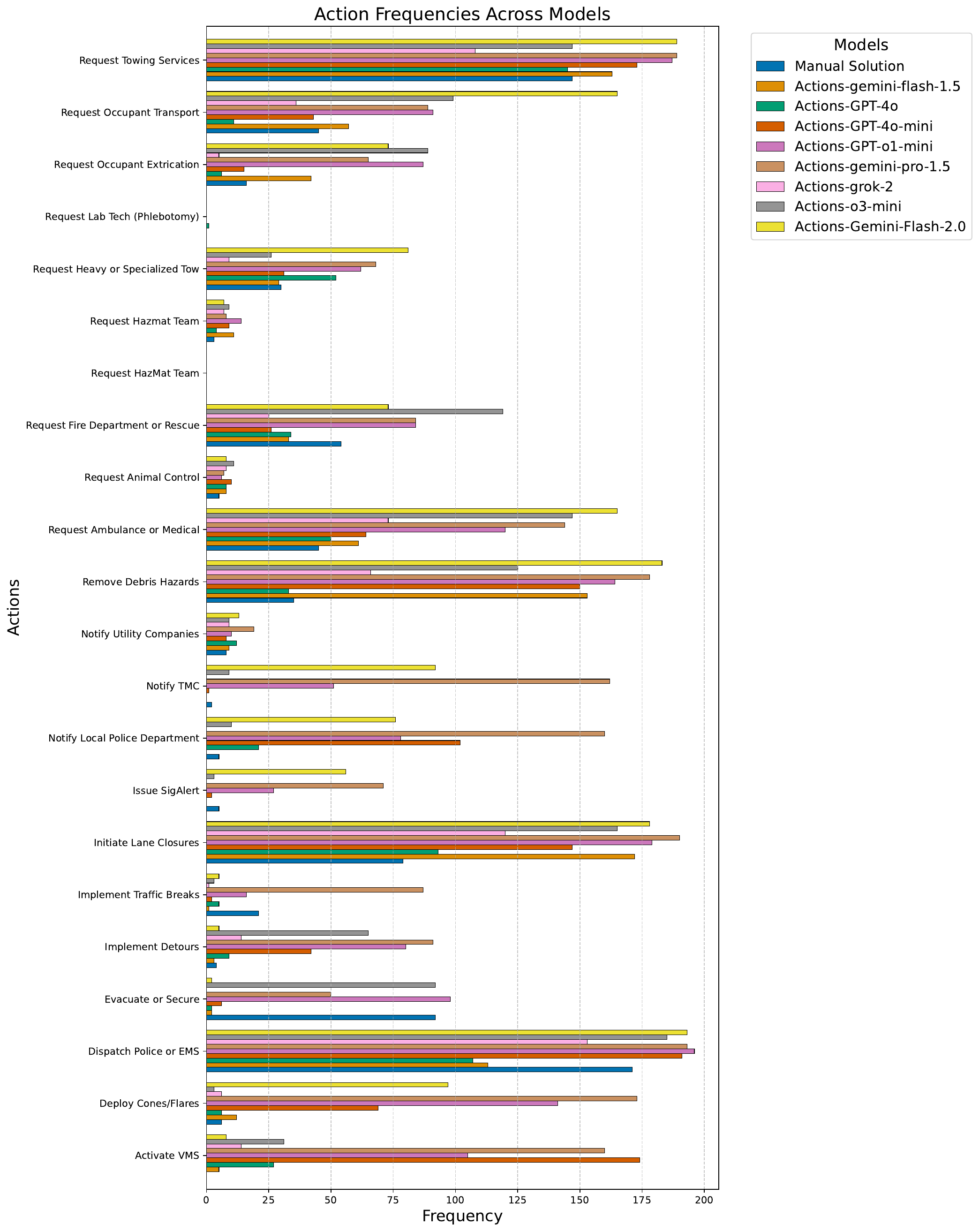}
    \caption{Action Frequency by Model}
    \label{fig:horiz}
\end{figure}

Figure~\ref{fig:horiz} shows the frequency of each action across all models after evaluation on 200 incident reports. Our analysis reveals several notable patterns in action frequency and model-specific behavior. In terms of action frequency patterns, actions such as “Dispatch Police or EMS” and “Request Towing Services” are consistently observed across all models, indicating strong agreement in these critical areas. Conversely, actions like “Activate VMS”, “Remove Debris Hazards”, and “Notify TMC” exhibit substantial variability, with noticeable differences in frequency between the models. Regarding model-specific patterns, Actions-gemini-pro-1.5 generally demonstrates a more aggressive approach by recommending actions more frequently overall, whereas Actions-grok-2 is more conservative, suggesting fewer actions despite its high similarity to the Manual Solution. Additionally, Actions-GPT-4o closely tracks the Manual Solution in key categories, even though its overall distribution differs. Notable discrepancies include the “Evacuate or Secure” action, which shows dramatic variation across models (Manual: 92, gemini-flash-1.5: 2, grok-2: 0, o3-mini: 92), and “Remove Debris Hazards”, which is recommended much more frequently by AI models than by the Manual Solution.

\subsection{Analysis of Hallucinations in LLMs During Plan Generation}

In the evaluation of large language models (LLMs), several types of hallucinations were observed across multiple model generations:

\begin{itemize}
    \item \textbf{Omission and invention of non-existing actions:} On ocassion, all the models generated a binary set of actions, either reducing the expected 21 actions (e.g. to 6) or inventing additional actions (e.g. produced an array of 32 actions), which highlights tendency of LLMs to hallucinate, omit and invent actions that were not specified, ignoring prompt constraints.
    
    \item \textbf{Overreaction:} The Gemini 1.5 Pro model consistently overreacted by including all available actions in its response, indicating impaired reasoning abilities. It lacked the discernment to filter unnecessary actions, which may suggest excessive conservatism in action selection.


    \item \textbf{Formatting Variability and Indirect Response:} All the models on occasion produced responses in different markup languages: Markdown, LaTeX, Python code and plain text - demonstrating inconsistencies in output format. Instead of generating a direct binary array, in multiple cases the models produced Python scripts for generating the binary action array instead of providing the exact binary array.

    \item \textbf{Placement of Reasoning:} In multiple instances, models placed reasoning for their action choices mixed with the binary array, rather than delivering a final response in the form of binary array only.. In this case, it's necessary to use a structured output capability, re-evaluate the prompt until correct format or apply the second prompting pass to correctly extract it.

    \item \textbf{Randomness in responses:} All the models tend to generate different response every time the prompt is evaluated. These variations may contain any of the mentioned previously issues including anomalous inclusion of not mentioned actions.

\end{itemize}

Overall, the variability in responses, response format and structure necessitates additional prompting and processing steps to accurately extract the binary action array. The prompt may also need the re-evaluation if the response does not contain the necessary answer.

\section*{Conclusion}

This study underscores the transformative potential of generative AI in revolutionizing traffic incident management. Our comparative analysis highlights significant disparities among language models, with some models demonstrating superior accuracy by closely replicating expert response plans.

Our comparative analysis of several state-of-the-art LLM architectures reveals significant disparities in their ability to generate these binary action vectors. The Hamming distance, within this analytical paradigm, provides an objective and interpretable measure of the semantic divergence between LLM-generated and expert-defined plans at the fundamental unit of operational response - the individual action selection.


The potential advantages afforded by the application of advanced LLMs for the rapid and accurate generation of well-structured binary action vectors in traffic incident management are substantial. The demonstrated capacity of these models to efficiently synthesize complex, context-aware response strategies into a clear, unambiguous, and directly executable binary format holds significant promise for achieving tangible improvements in roadway safety. However, our analysis reveals notable differences in the accuracy and effectiveness of various language models when generating response plans. Dense non-resoning models like GPT-4o and Grok 2 consistently outperform other models, generating plans that closely align with human-proposed solutions. 

We acknowledge that deploying LLMs in traffic management presents challenges such as potential data and guideline bias (since the inherent data-driven approach in model training) and ethical considerations surrounding automated decision-making regarding safety.

Despite these challenges, the potential benefits of AI-driven traffic management cannot be overstated. The ability of these models to swiftly generate complex, adaptable, and transparent (e.g. via chain-of-thought process) response plans has potential for far-reaching implications for improving safety, reducing operational costs, faster mitigation of incident impact on traffic flow and reduction of the economic and environmental impact of road incidents. This research serves as a crucial step toward realizing the full potential of AI in automated management of incidents.

Future research will focus on exploring the integration of Retrieval Augmented Generation (RAG) techniques to ground LLM responses in official incident response guidelines, investigating methodologies for automated guideline synthesis to align response plans with local regulations, and refining prompt design strategies to further optimize the accuracy of generated binary action vectors.

\bibliography{bib}

\begin{thebibliography}{10}
\expandafter\ifx\csname url\endcsname\relax
  \def\url#1{\texttt{#1}}\fi
\expandafter\ifx\csname urlprefix\endcsname\relax\def\urlprefix{URL }\fi
\expandafter\ifx\csname href\endcsname\relax
  \def\href#1#2{#2} \def\path#1{#1}\fi

\bibitem{cong}
A.~S. Pillai, Traffic management: Implementing ai to optimize traffic flow and
  reduce congestion, SSRN Electronic Journal (01 2024).
\newblock \href {https://doi.org/10.2139/ssrn.4916398}
  {\path{doi:10.2139/ssrn.4916398}}.

\bibitem{de2017traffic}
A.~M. De~Souza, C.~A. Brennand, R.~S. Yokoyama, E.~A. Donato, E.~R. Madeira,
  L.~A. Villas, Traffic management systems: A classification, review,
  challenges, and future perspectives, International Journal of Distributed
  Sensor Networks 13~(4) (2017) 1550147716683612.

\bibitem{s16020157}
K.~Nellore, G.~P. Hancke, \href{https://www.mdpi.com/1424-8220/16/2/157}{A
  survey on urban traffic management system using wireless sensor networks},
  Sensors 16~(2) (2016).
\newblock \href {https://doi.org/10.3390/s16020157}
  {\path{doi:10.3390/s16020157}}.
\newline\urlprefix\url{https://www.mdpi.com/1424-8220/16/2/157}

\bibitem{mah}
D.~Mahmud, H.~Hajmohamed, S.~Almentheri, S.~Alqaydi, L.~Aldhaheri, R.~A.
  Khalil, N.~Saeed, Integrating llms with its: Recent advances, potentials,
  challenges, and future directions, IEEE Transactions on Intelligent
  Transportation Systems PP (2025) 1--36.
\newblock \href {https://doi.org/10.1109/TITS.2025.3528116}
  {\path{doi:10.1109/TITS.2025.3528116}}.

\bibitem{xue2023}
Z.~Xue, C.~Xu, X.~Xu, Application of chatgpt in natural disaster prevention and
  reduction, Natural Hazards Research 3 (2023) 556--562.

\bibitem{goecks2023}
V.~G. Goecks, N.~R. Waytowich, Disasterresponsegpt: Large language models for
  accelerated plan of action development in disaster response scenarios,
  arXiv:2306.17271v1 [cs.LG] (2023).

\bibitem{ma2014automated}
Y.~Ma, W.~Zhang, J.~Lu, L.~Yuan, Automated generation of traffic incident
  response plans, Discrete Dynamics in Nature and Society 2014 (2014) 1--8.

\bibitem{ravish2021intelligent}
R.~Ravish, S.~R. Swamy, Intelligent traffic management: A review of challenges,
  solutions, and future perspectives, Transport and Telecommunication 22~(2)
  (2021) 1--14.

\bibitem{huang2007}
B.~Huang, X.~Pan, Integration of gis, traffic simulation, and optimization for
  incident response, Computers, Environment and Urban Systems 31 (2007).

\bibitem{zhang2024}
S.~Zhang, D.~Fu, et~al., Trafficgpt: Viewing, processing and interacting with
  traffic foundation models, Transport Policy 150 (2024) 95--105.

\bibitem{huang2007integration}
B.~Huang, X.~Pan, Integration of gis, traffic simulation, and optimization for
  incident response, Computers, Environment and Urban Systems 31 (2007).
\newblock \href {https://doi.org/10.1016/j.compenvurbsys.2006.06.001}
  {\path{doi:10.1016/j.compenvurbsys.2006.06.001}}.

\bibitem{salvagno2023artificial}
M.~Salvagno, F.~S. Taccone, A.~G. Gerli, Artificial intelligence
  hallucinations, Critical Care 27~(1) (2023) 180.

\bibitem{sun2024ai}
Y.~Sun, D.~Sheng, Z.~Zhou, Y.~Wu, Ai hallucination: towards a comprehensive
  classification of distorted information in artificial intelligence-generated
  content, Humanities and Social Sciences Communications 11~(1) (2024) 1--14.

\bibitem{sahoo2024addressing}
N.~R. Sahoo, A.~Saxena, K.~Maharaj, A.~A. Ahmad, A.~Mishra, P.~Bhattacharyya,
  Addressing bias and hallucination in large language models, in: Proceedings
  of the 2024 Joint International Conference on Computational Linguistics,
  Language Resources and Evaluation (LREC-COLING 2024): Tutorial Summaries,
  2024, pp. 73--79.

\bibitem{mukhopadhyay2023emergency}
e.~a. Mukhopadhyay, Dynamic models for emergency response: Bridging theory and
  practice, Journal of Emergency Management (2023).

\bibitem{10737408}
T.-H. Vu, S.~Kumar~Jagatheesaperumal, M.-D. Nguyen, N.~Van~Huynh, S.~Kim, Q.-V.
  Pham, Applications of generative ai (gai) for mobile and wireless networking:
  A survey, IEEE Internet of Things Journal 12~(2) (2025) 1266--1290.
\newblock \href {https://doi.org/10.1109/JIOT.2024.3487627}
  {\path{doi:10.1109/JIOT.2024.3487627}}.

\bibitem{zhou2024dynamicroutegpt}
Z.~Zhou, B.~Zhou, H.~Liu,
  \href{https://arxiv.org/abs/2408.14185}{Dynamicroutegpt: A real-time
  multi-vehicle dynamic navigation framework based on large language models},
  arXiv preprint arXiv:2408.14185 (2024).
\newline\urlprefix\url{https://arxiv.org/abs/2408.14185}

\bibitem{tao2023}
X.~Tao, Exploring trustworthiness issues about disaster-related information
  generated by artificial intelligence (2023).

\bibitem{zheng2023chatgpt}
O.~Zheng, M.~Abdel-Aty, D.~Wang, Z.~Wang, S.~Ding, Chatgpt is on the horizon:
  Could a large language model be all we need for intelligent transportation?,
  arXiv preprint arXiv:2303.05382 (2023).

\bibitem{LAI1994486}
Y.-J. Lai, T.-Y. Liu, C.-L. Hwang,
  \href{https://www.sciencedirect.com/science/article/pii/0377221794902828}{Topsis
  for modm}, European Journal of Operational Research 76~(3) (1994) 486--500,
  facility Location Models for Distribution Planning.
\newblock \href {https://doi.org/https://doi.org/10.1016/0377-2217(94)90282-8}
  {\path{doi:https://doi.org/10.1016/0377-2217(94)90282-8}}.
\newline\urlprefix\url{https://www.sciencedirect.com/science/article/pii/0377221794902828}

\bibitem{zavadskas2016development}
E.~K. Zavadskas, A.~Mardani, Z.~Turskis, A.~Jusoh, K.~M. Nor, Development of
  topsis method to solve complicated decision-making problems—an overview on
  developments from 2000 to 2015, International journal of information
  technology \& decision making 15~(03) (2016) 645--682.

\bibitem{akilan2018effect}
T.~Akilan, Q.~J. Wu, H.~Zhang, Effect of fusing features from multiple dcnn
  architectures in image classification, IET Image Processing 12~(7) (2018)
  1102--1110.

\bibitem{boulahia2021early}
S.~Y. Boulahia, A.~Amamra, M.~R. Madi, S.~Daikh, Early, intermediate and late
  fusion strategies for robust deep learning-based multimodal action
  recognition, Machine Vision and Applications 32~(6) (2021) 121.

\end{thebibliography}

\end{document}